\documentclass[conference]{IEEEtran}

\usepackage{cite}
\usepackage{amsmath,amssymb,amsfonts}
\usepackage{algorithmic}
\usepackage{graphicx}
\usepackage{textcomp}
\usepackage{xcolor}
\usepackage{amsthm}
\usepackage{url}
\def\BibTeX{{\rm B\kern-.05em{\sc i\kern-.025em b}\kern-.08em
    T\kern-.1667em\lower.7ex\hbox{E}\kern-.125emX}}

\theoremstyle{definition}
\newtheorem{theorem}{Theorem}
\newtheorem{algorithm}{Algorithm}
\newtheorem{problem}{Problem}
\newtheorem{definition}[theorem]{Definition}
\usepackage{xspace}
\usepackage{comment}
\newcommand\gzip{\texttt{gzip}\xspace}
\newcommand\gunzip{\texttt{gunzip}\xspace}
\newcommand\libdeflate{\texttt{libdeflate}\xspace}

\newcommand\pugz{\texttt{pugz}\xspace} 

\begin{document}

\title{Parallel decompression of \gzip-compressed files and random access to DNA sequences}


\author{\IEEEauthorblockN{Ma\"el Kerbiriou}
\IEEEauthorblockA{INRIA Lille - Nord Europe \\
\textit{CRIStAL, Universit\'e de Lille} \\
Lille, France\\
mael.kerbiriou@inria.fr}
\and
\IEEEauthorblockN{Rayan Chikhi}
\IEEEauthorblockA{C3BI USR 3756 \\
\textit{Institut Pasteur \& CNRS}\\
Paris, France \\
rayan.chikhi@pasteur.fr}
}

\maketitle

\begin{abstract}
Decompressing a file made by the \texttt{gzip} program at an arbitrary location is in principle impossible, due to the nature of the DEFLATE compression algorithm. Consequently, no existing program can take advantage of parallelism to rapidly decompress large  \texttt{gzip}-compressed files. This is an unsatisfactory bottleneck, especially for the analysis of large sequencing data experiments.
Here we propose a parallel algorithm and an implementation, \pugz, that performs fast and exact decompression of any text file. 
We show that \pugz is an order of magnitude faster than \texttt{gunzip}, and 5x faster than a highly-optimized sequential implementation (\texttt{libdeflate}).
We also study the related problem of random access to compressed data. We give simple models and experimental results that shed light on the structure of \gzip-compressed files containing DNA sequences.
Preliminary results show that random access to sequences within a \gzip-compressed FASTQ file is almost always feasible at low compression levels, yet is approximate at higher compression levels.
\end{abstract}


\section{Introduction}
The compression of genomic sequencing data continues to be a busy research area, using fundamental techniques such as Lempel-Ziv parsing~\cite{lz77}, Burrows-Wheeler transform~\cite{bwt}, or de Bruijn graphs~\cite{leon}. Despite many promising advances, the most popular and practical program for compressing raw sequencing data nowadays remains the Lempel-Ziv based \gzip program, arguably due to its speed and availability on all systems. There exist parallel programs for speeding-up \gzip compression, e.g. \texttt{pigz}\footnote{\url{github.com/madler/pigz}}. 
The underlying compression algorithm of \gzip, DEFLATE, easily lends itself to processing of blocks of data concurrently. 

However, the decompression aspects are more challenging and have been less studied, leaving us with only sequential algorithms. Investigating decompression would not be a pressing matter if it was already an IO-bound operation, but actually only around 30-50 MB of compressed data per second can be processed by \texttt{gunzip} on a modern processor. This is much below the 500 (resp. 100--200) MB/sec read throughput of SATA solid-state (resp. mechanical) drives. In the near future, the wide availability of NVMe solid-state drives will allow for even faster reads, up to 3 GB/sec. Therefore a potential slowdown of 1-2 orders of magnitudes exists at the beginning of many tools, and in particular bioinformatics pipelines that take compressed files as input. 

If one could efficiently perform random accesses to content in gzipped files, it would unlock multi-core decompression. As a result, data would be streamed significantly faster than if it was read from an uncompressed file. However, the DEFLATE compression algorithm in \gzip works in such a way that accessing a random location does in principle require to read the whole file up to the desired location. This is because (1) variable-length blocks are created without any index nor explicit block starts/ends, and (2) back-references are made to strings that appear previously in the decompressed file, which is the essence of Lempel-Ziv parsing. A more precise description of DEFLATE is provided later in the article. 

To build some intuition on why random access is hard, let an example file contain $n$ identical repetitions of a sentence then followed by $m$ identical repetitions of a completely different sentence. Suppose that neither the sentences lengths, $n$ nor $m$ are known. The compression algorithm in \gzip uses heuristics yet is very likely to always encode repetitions of the sentences using back-references and not in clear form, except at two locations: (i) the beginning of the file, and (ii) the DEFLATE block(s) where the second sentence appears for the first time. A random access towards the end of the file would require to somehow obtain information that is present only at location (ii). Yet, since this location cannot be easily guessed, it is challenging to design a better strategy than to fully decompress the file.

There exist techniques to create so-called "blocked gzip" files, i.e. files that are compressed as a sequence of small and independent blocks. Such files can also contain an additional index that enables random access and parallel decompression. However as we will see in Section 2, these techniques are i) not so wide-spread and ii) yield worse compression ratios.

In this work, we will focus on the parallel decompression of files compressed by the \gzip program. In bioinformatics, virtually every tool that processes large amounts of raw sequencing data begins by reading large \texttt{.fastq.gz} file(s).
We first present simple models to provide a general understanding of \gzip compression. Parallel decompression would be straightforward if one could perform decompression at random locations. We therefore start by studying the feasibility of random access in compressed files, with a focus on files containing DNA sequences. Random access turns out to be challenging for files with high compression ratio. We thus propose an alternative strategy via the design of a new, general-purpose, exact and parallel \gzip decompression algorithm that can process any arbitrary ASCII file. 
We demonstrate the usefulness of this algorithm by showing significant speed-ups compared to the baseline method (gunzip) and a highly optimized implementation (libdeflate).

\section{Related works}

Several algorithms have been proposed for the compression of genomic data (for a survey, see ~\cite{deorowicz2013data}). Yet, the most popular solution arguably remains \gzip. It achieves roughly a 3-fold size reduction on files in the FASTQ format. While files produced by \gzip are not designed to make random access efficient, there exists variants that permit this operation. For instance, a modification of Lempel-Ziv parsing has been proposed to support random access at the expense of slightly lower compression ratio~\cite{Kreft2010}, but the \gzip program does not implement it. 
In the same spirit, the \texttt{tabix}/\texttt{bgzip} programs from the SAMtools/HTSlib project~\cite{samtools} create so-called blocked files that are indexed and \gzip-compatible\footnote{\url{blastedbio.blogspot.fr/2011/11/bgzf-blocked-bigger-better-gzip.html}}.
One can also create an index of blocks within a classical \gzip-compressed file for later faster access to random locations~\cite{li_2014}.
The index is created during an initial sequential decompression of the whole file. This essentially solves random access, except that the technique is not so widespread as it requires a separate file or a different file format, and does not apply when one only needs to read a given compressed file once.
Anecdotally, a majority of compressed files hosted in the Sequence Read Archive repository and uploaded in 2018 are not compressed in blocks.

Previous efforts have aimed to improve compression via better algorithms or hardware acceleration, e.g. in~\cite{sadakane2000improving,abdelfattah2014gzip}.
Efficient decompression has been relatively less studied, except in the following works. A fine-grained hardware-accelerated decompression algorithm achieves a 2x speedup compared to \texttt{gunzip}~\cite{sitaridi2016massively}. A recent unpublished highly-optimized sequential implementation (\texttt{libdeflate}) achieves roughly a 3x speedup\footnote{\url{github.com/ebiggers/libdeflate}}.
%
Perhaps the closest work to ours is an attempt to recover corrupted DEFLATE-compressed files~\cite{brown2011reconstructing,brown2013improved}. It is proposed that damaged \gzip-compressed files containing English text can be partially reconstructed from random locations. However, the bulk of the method hinges on frequency analysis which does not directly translate to files containing DNA, and also settles for lossy reconstruction.

\section{DEFLATE algorithm}

A file compressed using \gzip (or \texttt{zlib}) consists of a header, a sequence of bytes compressed using the DEFLATE algorithm, and a final checksum~\cite{RFC1950}. In this section we provide a high-level description of DEFLATE, where many details will be omitted~\cite{RFC1951}.
Input data is partitioned into blocks, that may or may not be compressed, depending on whether the data is ccompressible. To simplify the exposition we will consider (in this section only) the creation of a single compressed block, which is performed in two stages. The first stage is LZ77-style parsing, which encodes the input data into a succession of literals (i.e. unmodified bytes) and offset/length integer pairs, later denoted as \emph{matches}, referring to previous occurrences of decompressed characters. In the original definition of LZ77~\cite{lz77}, the parsing has to be of the form "match-literal-match-literal-\ldots", as follows.

\begin{definition}[LZ77 parsing] 
Let $s$ be a string. A LZ77 parsing of $s$ is a sequence of phrases $p_1,\ldots,p_z$, where each $p_i$ is a tuple consisting of two integers (offset and length) and a character. The first phrase $p_1$ encodes an empty match $(0,0)$ along with the first character of $s$. If $p_1,\ldots,p_{i-1}$ represent $s$ up to position $j$, then $p_i$ records the offset and length of the longest prefix of $s[j+1\ldots]$ that occurs in $s[\ldots j]$, along with the character that immediately follows the prefix.
\end{definition}

We now define a more general flavor that better abstracts the product of DEFLATE. 

\begin{definition}[mixed LZ77-style parsing] 
Let $s$ be a string, a mixed LZ77-style parsing (mLZ) of $s$ is a sequence $p'_1,\ldots,p'_z$, where each $p'_i$ is either a single character of $s$ or a pair of offset/length integers that refers to a substring of $s$ within  $p'_1,\ldots,p'_{i-1}$.
\end{definition}

LZ77 parsing can be seen as a special case of mLZ.  
In DEFLATE, matches are of length between 3 and 258 bytes, and are made within a sliding window of length 32 KB, i.e. offsets are never larger than 32 KB. 

The second stage of DEFLATE uses Huffman coding to entropy-compress the output of the first stage. In practice, DEFLATE uses two Huffman code trees, but going into more details will not be necessary here. An important point however is that Huffman compression is reset at the beginning of each block. To summarize, we can abstract the compression process as follows.

\begin{algorithm}[DEFLATE compression]
\label{alg:comp}
Let $s$ be a string, DEFLATE compression produces a Huffman-compressed representation $C$ from a mixed LZ77-style parsing of $s$.
\end{algorithm}

Note that Algorithm~\ref{alg:comp} is not fully specified, as there are multiple ways to encode a string as a mLZ (e.g. using more literals at the expense of longer matches, or the opposite). Decompression, however, is more straightforward.

DEFLATE decompression processes a stream of Huffman-coded symbols, and writes decoded bytes into a circular buffer of fixed size $W=32$ KB, that we call \textbf{\emph{context}}. A data byte is simply appended to the buffer if it corresponds to a literal value, otherwise a match is decoded by copying a certain number of bytes from a previous position in the buffer. 

\begin{algorithm}[DEFLATE decompression]
\label{algo:dec}
The algorithm $\texttt{decompress}(C,w)$ takes as input a representation of an entropy-compressed mixed LZ77-style parsing $C$, and a circular buffer $w$ (\emph{context}). While parsing $C$ from left to right, decoded literals are sequentially appended to $w$, and matches are copied from previous positions in $w$.
\end{algorithm}

In essence, \gzip performs decompression by calling $\texttt{decompress}(C,w)$ on a stream of compressed data, using an initially empty context $w$. Clearly the first phrase of $C$ needs to be a literal. And at the beginning of $C$, matches can only copy characters from what was previously appended to the context. 

\section{Feasibility of random access}

\subsection{Problem statement}

Random access could be performed into positions in the original uncompressed file, or positions within the compressed stream. In this article we will exclusively focus on the latter.
Performing decompression at a random location in a DEFLATE stream requires the knowledge of two pieces of information: the start position of a nearby block, and a fully decompressed 32 KB context.
We will later see that a block position can be reliably guessed using exhaustive search. Obtaining a decompressed context is the challenging part. We formulate the problem as follows:

\begin{problem}[Random access in DEFLATE stream]
\label{prob:random}
Given a mixed LZ77-style parsing $C$ corresponding to the compression of an unknown string $s$, return the smallest possible integer $i$ and a context $w$ such that $\texttt{decompress}(C[i\ldots],w)$ returns a suffix of $s$.
\end{problem}

Here, the input $C$ can be thought as a suffix of a mLZ produced by Definition~\ref{alg:comp}. We allow a prefix of $C$ to be discarded, typically to avoid having to decompress any back-reference to the (unknown) context that precedes $C$. Note that the problem does not appear to be well-formulated, as checking that the decoded string is indeed a suffix of $s$ is not possible without the knowledge of $s$. However, one can make sure that there is no ambiguity as to how the output could have been decompressed.

\subsection{Undetermined context propagation technique \label{sec:technique}}

When starting decompression at a random location, reliably guessing the contents of the context $w$ that precedes the decompression location does not appear to be a trivial task. Therefore it makes sense to start with an \emph{undetermined} context, e.g. a context that consists only of question mark '?' characters, that we will call undetermined characters.

\begin{figure*}[!htb]
    \centering
    \includegraphics[width=.8\textwidth]{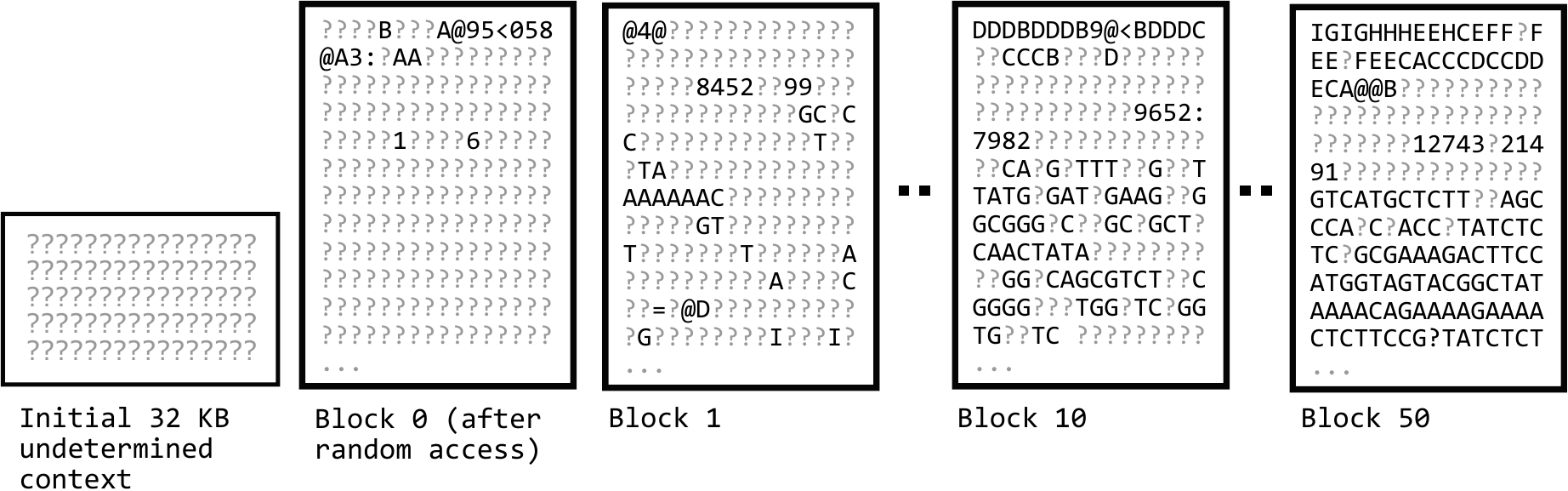}
\caption{Illustration of the decompression process starting from a random location in a \gzip-compressed FASTQ file. The location of the first following block is determined. A 32 KB context is initialized, consisting of undetermined '?' characters. Then, all subsequent blocks are decompressed into literals and back-references. Each back-reference copies characters from its 32 KB context. The first 192 bytes of a selection of compressed blocks are shown, on a real Illumina FASTQ file. Subsequent blocks contain less and less undetermined characters, as more literals are decompressed and back-referenced.}
\label{fig:blocks}
\end{figure*}

We postulate that for certain compressed files, running the $\texttt{decompress}$ procedure for a certain amount of time ends up with decoded contexts that no longer contain undetermined characters.
 On the one hand, back-references (matches) to initially undetermined characters propagate those characters into later contexts. On the other hand, characters that are compressed as literal values contribute to making the current context \emph{less} undetermined, and are possibly propagated into future contexts as well. Thus if sufficiently many literals are produced by the compression algorithm, there is a chance that any undetermined character no longer get back-referenced after a sufficiently long decompression.
 An illustration of the process is given in Figure~\ref{fig:blocks}.

\subsection{Experimental test on random DNA \label{sec:expval}}

To further study the postulate above, 
we experimentally created a random DNA string of length 1 Mbp and compressed it with \gzip at various compression levels. For each level we ran \texttt{decompress} starting from the second compressed block, using a fully undetermined context. At the default compression level (\texttt{-6}), the average offset of matches is $o_a=3602$. We then counted the number of undetermined decompressed characters in non-overlapping windows of size $o_a$. While \gzip uses a sliding window for back-references, using non-overlapping windows more appropriately mimics the model developed later in Section~\ref{sec:greedmodel}. 

\begin{figure}[!htb]
\centering
\includegraphics[width=.9\columnwidth]{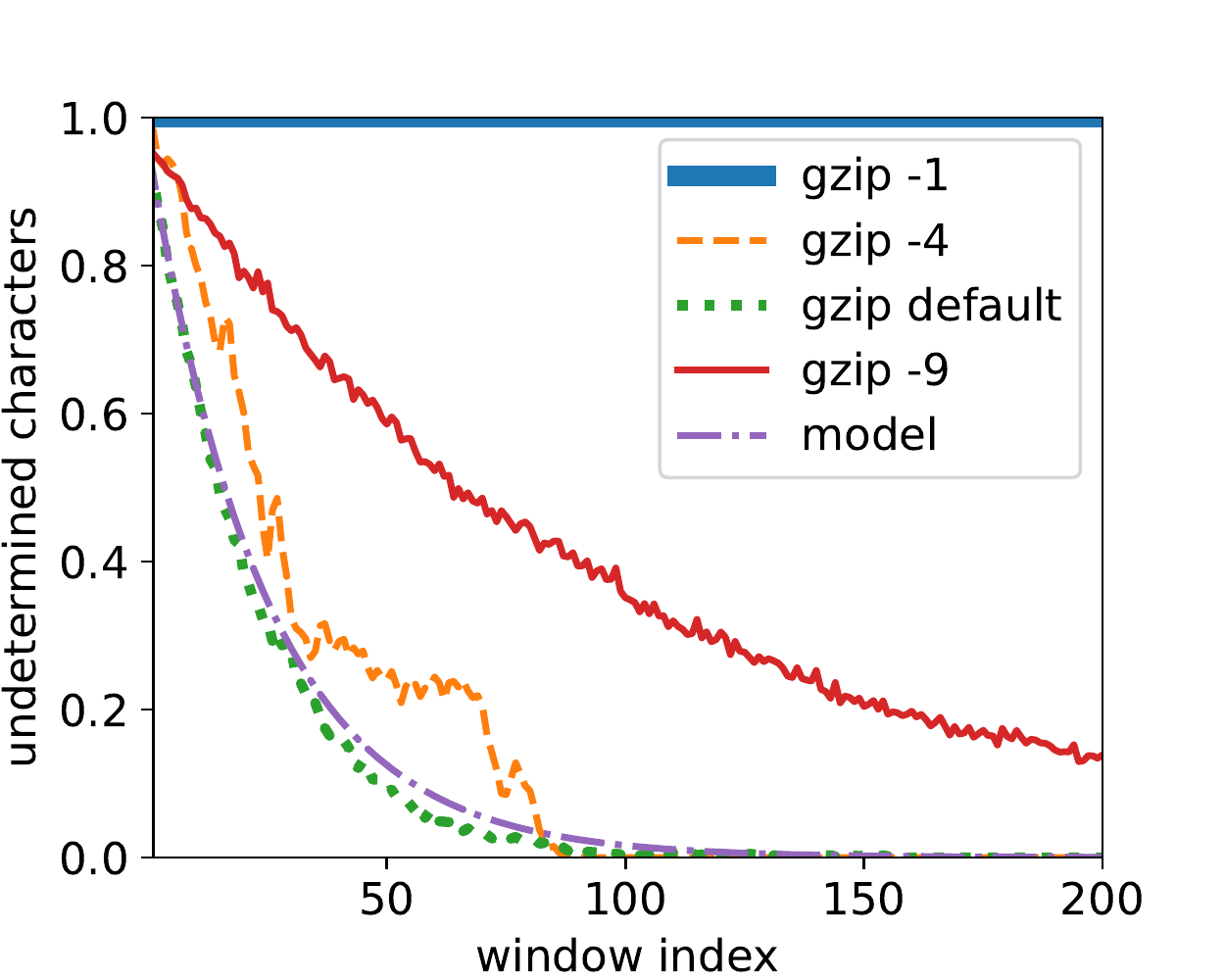}
\includegraphics[width=.9\columnwidth]{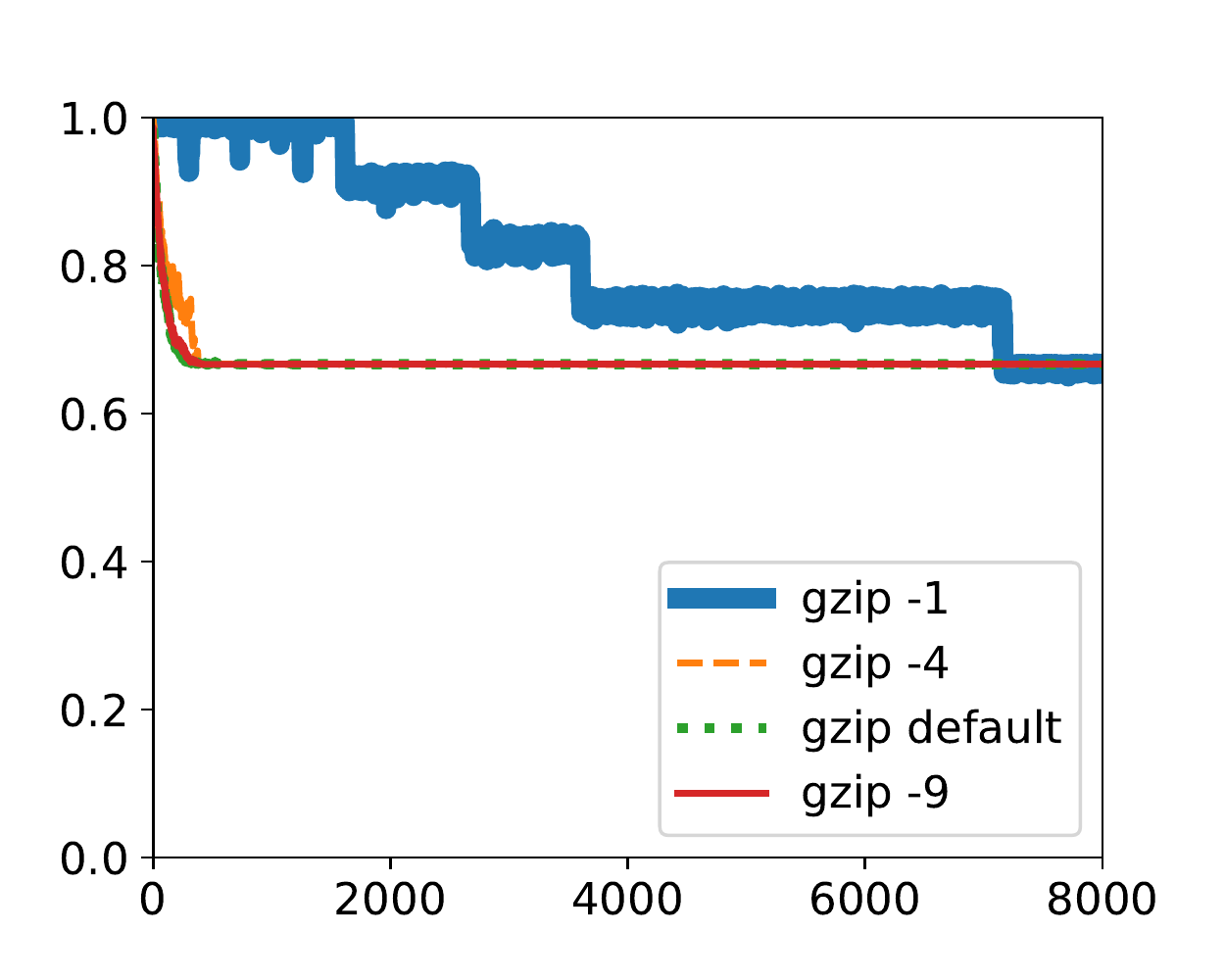}
\caption{Counting undetermined characters in the decompression stream of \gzip-compressed files: random DNA (top) and FASTQ-like with random DNA sequences (bottom). Files were decompressed from block 2 starting with an undetermined context. In the first four lines of the legend, the $x$ axis corresponds to indices of non-overlapping windows of average offset length $o_a$. The $y$ axis corresponds to the percentage of undetermined characters in a window. The 'model' line corresponds to our idealized non-greedy parsing model developed in Section~\ref{sec:greedmodel}: here the $x$ axis corresponds to indices $i$ and the $y$ axis are the values $(1-L_i)$.}
\label{fig:model}
\end{figure}

We refer the reader to the top part of Figure~\ref{fig:model}, disregarding for now the "model" line.
At the default and \texttt{-4} compression levels, after decompressing around 150*3600=540 KB of random DNA compressed data, the proportion of undetermined characters in a window vanishes. At compression level \texttt{-9}, the percentage of undetermined characters eventually vanishes after window index 790 (not shown in Figure~\ref{fig:model}). If no undetermined character remains, what is decompressed must be a correct suffix. Thus,  this experiment hints that the approach proposed in Section~\ref{sec:technique} provides a solution to Problem~\ref{prob:random} on files containing random DNA that were compressed with level \texttt{-4} and higher.

It may appear intriguing that at the lowest compression levels (\texttt{-1} to \texttt{-3}), the whole file starting from block 2 is encoded using only matches and zero literals and thus solving Problem~\ref{prob:random} appears to be impossible. In fact this phenomenon occurs also on much larger files (200 MB) containing random DNA. However we will later see that the model proposed in Section~\ref{sec:simpmodel} explains this effect, and that it is not a critical problem as we are mainly interested in FASTQ files. Indeed, FASTQ files have a different structure than files containing random DNA and their compression using \gzip behaves differently, as we see next.

\subsection{Compression of FASTQ-like strings}


The FASTQ format is a text file format that consists of groups of four lines: a header, a DNA sequence, another header, and a string of ASCII-encoded quality values.  Since DNA sequences are separated by headers and quality values, compression of FASTQ files yield larger offsets than in files containing only random DNA (i.e. as in Section ~\ref{sec:expval}).
This will have an impact on compression as, at least at the lowest compression levels, \gzip favor matches with small offsets and may output literals instead of matches with large offsets.

We created a FASTQ-like string of length 150 MB by repeating 150 random DNA characters followed by 300 'x' characters. Figure~\ref{fig:model} (bottom) shows that, at the lowest compression level, all DNA characters eventually become decompressed as literals, but only after around 25 MB ($>7000o_a$) of decompressed data.
Consequently, random access in FASTQ-like files appears feasible at any compression level.

\section{\gzip compression of DNA sequences}

In this section, we provide simple models that explain our observations from Section~\ref{sec:expval}. Concretely, we will quantify the number of literals emitted during \gzip compression of DNA sequences.

\subsection{Simplified model for the compression of random DNA~\label{sec:simpmodel}}

Let us consider two blocks of random\footnote{A random DNA model may appear inapplicable to files containing short sequencing reads, but in fact previous studies have established that reads are difficult to compress~\cite{deorowicz2013data}, in particular with \gzip, making them behave similarly to random data. To check this, we extracted 32 KB windows of sequences positions 0, 1 MB and 20 MB of 10 Illumina datasets (from the Results section) and tested their randomness via compression~\cite{ryabko2005using}.
All windows except in 2 datasets showed compression ratios above 2.1 bits/character using \texttt{bzip2 -9}, i.e. above a naive 2-bit conversion and indicating that the files behave similarly to random sequences. The remaining windows in 2 datasets compressed to respectively 1.7 and 1.9 bits/character but the corresponding reads had low GC-content and adapter sequences, respectively.}
 DNA of length $W$ each, and let us compress the second block using only matches to the first one.
The DEFLATE algorithm would compress the second block using also matches within itself or matches overlapping both blocks, but separating matches between blocks will simplify the analysis. 
Under another simplifying assumption that each position has an independent chance to match, the probability that a match of length $k$ occurs at any given position in the second block is $$p_k=1-(1-\frac{1}{4^k})^{W-k+1}\approx 1-e^{-4^{-k}(W-k+1)}$$
by a Poisson approximation.
The probability that all positions in the second block have a match of length $k$ is $p_k^{W-k+1}$. For the minimum matching length in gzip ($k=3$) and its typical context size ($W=2^{15}$), $p_k \geq 1-10^{-225}$ and $p_k^{W-k+1} \geq 1-(W-k+1)10^{-225} \geq 1-10^{-220}$ by Bernoulli inequality. Therefore 
all positions in the second block have a match, with probability essentially 1.

This model hints that a DEFLATE-based algorithm such as \gzip could in principle encode the suffix of a random DNA string using only matches of length 3 or more, i.e. without emitting any literal after some point. In such a scenario, Problem~\ref{prob:random} would seem impossible to solve. This is indeed what we had observed in Section~\ref{sec:expval} when a random DNA string is encoded using \texttt{gzip -1} (lowest compression level).

\subsection{Non-greedy parsing}

Fortunately, \gzip attempts to maximize compression ratio and sometimes uses literals instead of matches. It can indeed be less costly to emit a couple literals, say $3$ literals encoded using $4$ bits each (12 bits total), as opposed to a short match that has a large offset, say $32000$, that would require around $16$ bits to be encoded in the worst case. While this effect would be interesting to model, we turn our attention to another related property of the DEFLATE algorithm called \emph{lazy matching}~\cite{RFC1951} or \emph{non-greedy parsing} (as analyzed in~\cite{nongreedy}). This is a strategy that improves compression by favoring longer matches. 

\begin{algorithm}[non-greedy parsing~\cite{RFC1951}]
The mLZ of a string $s$ is constructed sequentially, by induction, considering the following cases. If the maximal length of a match at position $i$ is $l$, but the maximal length of a match at position $i+1$ is $l_2 > l$, then emit literal $s[i]$ followed by one of the longest match(es) at position $i+1$ and advance in $s$ by $l_2$ positions. If not, emit a match at position $i$ if one exists (and advance by $l$ positions), otherwise emit the literal $s[i]$ and advance by one position. 
\end{algorithm}

For DNA sequences, as we will see next, non-greedy parsing turns out to greatly increase the number of literals being emitted during the compression. In \gzip, it is always used except in the three lowest compression modes (flags \texttt{-1}, \texttt{-2}, or \texttt{-3}). 

\subsection{Compression of random DNA under non-greedy parsing \label{sec:greedmodel}}

Keeping our previous compression model of two blocks of random DNA, along with the assumption that matches occur independently, let $p^\ell$ be the probability of a literal being emitted at a given position under non-greedy parsing. Using a decomposition into disjoint events depending on the length of the first match, $$p^\ell=\sum_{k\geq 3} p_k(1-p_{k+1})p_{k+1}$$ 
where $p_k(1-p_{k+1})$ is the probability that the current position has a maximal match of length $k$, and $p_{k+1}$ the probability that a match of length at least $(k+1)$ occurs at the next position. Given an average match length $l_a$, an experimentally-verified approximation for the number of literals being emitted due to non-greedy parsing is $E^\ell=p^\ell W/(l_a+2)$, guided by the intuition that on average only one position out of $l_a+1$ will be available for matching, and non-greedy parsing adds one literal (hence the term $l_a+2$). For $W=2^{15}$ and $l_a$ experimentally determined to be $7.6$, 
$E^\ell\approx 1283$. Thus in our model, around $E^\ell/W=4\%$ of characters would be encoded in the second block as literals.

Currently our compression model has only two blocks. Consider an extension to $n$ blocks, still with matches restricted to within a block before the current one. How many literals or copies of literals would make up subsequent blocks? Let $L_i$ be percentage of literals or copies of literals at block $1\leq i\leq n$. Our previous estimations gave $L_1=E^\ell/W$. Suppose that any subsequent block is compressed using $E^\ell$ literals due to non-greedy parsing, and the remaining characters are independently sampled from the previous block. Then we obtain a classic arithmetic progression: 
\begin{align*} 
L_{i+1} &= \frac{E^\ell + (W-E^\ell)L_i}{W}\\
&=L_1+(1-L_1)L_i \\&= 1-(1-L_1)^{i+1}
\end{align*}
 Therefore under this simplified non-greedy parsing model in the compression of random DNA, the number of characters that are not copies of literals decreases exponentially. 

\subsection{Experimental validation}

We now refer the reader to the "model" line of the top part of Figure~\ref{fig:model}, which plots the proportion $(1-L_i)$ of undetermined characters remaining within a window, using $L_1=4\%$. We notice that this non-greedy parsing model fits reasonably well the actual behavior of gzip at the default compression level. 
As a side note, the maximum compression level (\texttt{-9}) produces matches of higher average offset ($o'_{a}=12755$) that also fit our non-greedy model (data not shown).






\section{Algorithm and implementation details}

So far, it appears possible to (i) extract suffixes from a random location in a \gzip-compressed FASTQ file, which would lead to (ii) decompress whole {\gzip}-compressed FASTQ files and possibly even arbitrary files efficiently in parallel. In this section, we describe two key components to achieve these goals: robust detection DEFLATE block start positions, and a heuristic algorithm that extracts DNA sequences from a decompressed DEFLATE block. And finally we will describe a general parallel decompression algorithm.

\subsection{Detection of DEFLATE block start positions \label{sec:block_pos}}

DEFLATE blocks that make up a \gzip-compressed file are neither indexed nor aligned on byte boundaries. Therefore they can occur at any bit offset. Once a block start position is known, decompressing the block gives the bit position of the next block. Hence once the position of an initial block is determined, the rest of the file can be decompressed. Our strategy consists in trying to run the decompression routine from every bit position. Each time the decompression of a putative block fails, we backtrack to the next bit that follows the start position of the failed block. From a performance standpoint, it is important to fail early and as quickly as possible. We implemented a set of stringent checks (see Appendix~\ref{appendix:block}), and used branch probability hints to guide the C++ compiler.

Once a block decompression succeeds, we decompress five more blocks to confirm that we are effectively synced to the DEFLATE stream. A failure to do so would backtrack to the bit after the first decompressed block. Overall this strategy robustly finds the next block start position after a certain bit offset, in around 100--300 milliseconds. 

\subsection{Heuristic determination of sequence-resolved blocks \label{sec:determine}}

We provide a prototype software implementation that performs random access to DNA sequences in a compressed FASTQ file: \url{https://github.com/rchikhi/fqgz}.
The implementation uses a heuristic linear-time procedure to extract DNA-like segments from a decompressed FASTQ file block that still contains undetermined characters. For details, see Appendix~\ref{appendix:heuristic}. 

A \gzip-compressed block is said to be \emph{sequence-resolved} if the parser returns sufficiently many sequences (determined by a fixed threshold), and none of these sequences contain an undetermined character. There may still be undetermined characters elsewhere in the block, i.e. in headers and/or quality values. We note also that our current implementation may, in pathological cases, (1) wrongly determine that a block is sequence-resolved or (2) fail to recognize that a block is sequence-resolved. Note also that subsequent blocks following a sequence-resolved block are not necessarily sequence-resolved.

In practice, none of the above problems are critical as the implementation will only be used to estimate, in Section~\ref{sec:res-succrate}, whether undetermined nucleotides remain in DNA sequences long after a random access is performed using an undetermined context. In its current state, the implementation is not a robust FASTQ parser. However, it is suitable for forensics applications, e.g. when dealing with data corruption in compressed FASTQ files.

As random access appears to be a challenging task, we now turn to a different strategy for whole-file decompression.

\subsection{Full file decompression in two passes (\pugz)}

To decompress a \gzip file in parallel, we propose an exact and general-purpose parallel decompression algorithm. It is based on breaking the compressed file at block boundaries, into $n \geq 2$ roughly equal parts ${C_1,\ldots,C_n}$.

So far we have considered that, in a random access setting, the initial context of a block should contain repetitions of the same undetermined character (as per Section~\ref{sec:technique}). We propose here a finer technique that consists in representing the initial context using a sequence of unique symbols $\hat{w}=[U_0, \ldots, U_{32767}]$, allowing to keep track of their propagation when back-referenced by matches in the decompressed stream.

This special context encoding can be leveraged to design a parallel decompression algorithm
in two passes. During the first pass, the \gzip-compressed file is decompressed in
parallel on $n\geq 2$ threads but no character is output. The origin of all back-referenced characters is recorded, and traced back to the initial context of the thread.

Formally, each thread $i \geq 0 $ runs $\texttt{decompress}(C_{i},\hat{w_{i}})$, yielding decompressed text $D_{i}$. 
Let $w_{i+1}$ be the last 32 KB characters of $D_i$. The idea is that symbols in $w_{i+1}$ can be used to resolve the initially unknown context $\hat{w_{i+1}}$ that was given to thread $i+1$. In particular, observe that thread 0 produces a suffix $w_1$ that does not contain any undetermined character. But in general, $w_{i+1}$ may contain undetermined characters for $i\geq 1$.
This is why we need a second pass, where for $i=2\ldots n$,
a sequential replacement of all undetermined characters in $w_{i}$ is performed using characters from $w_{i-1}$.
Then, after all the $w_i$ for $i\geq 1$ no longer contain any undetermined character, in parallel each thread $i\geq 1$ replaces occurrences of  symbol $U_j$ in $D_i$ by $w_{i}[j]$ for all $0\leq j\leq 32767$, yielding the fully decompressed file.

\begin{figure*}[!htb]
    \centering
    \includegraphics[width=.75\textwidth]{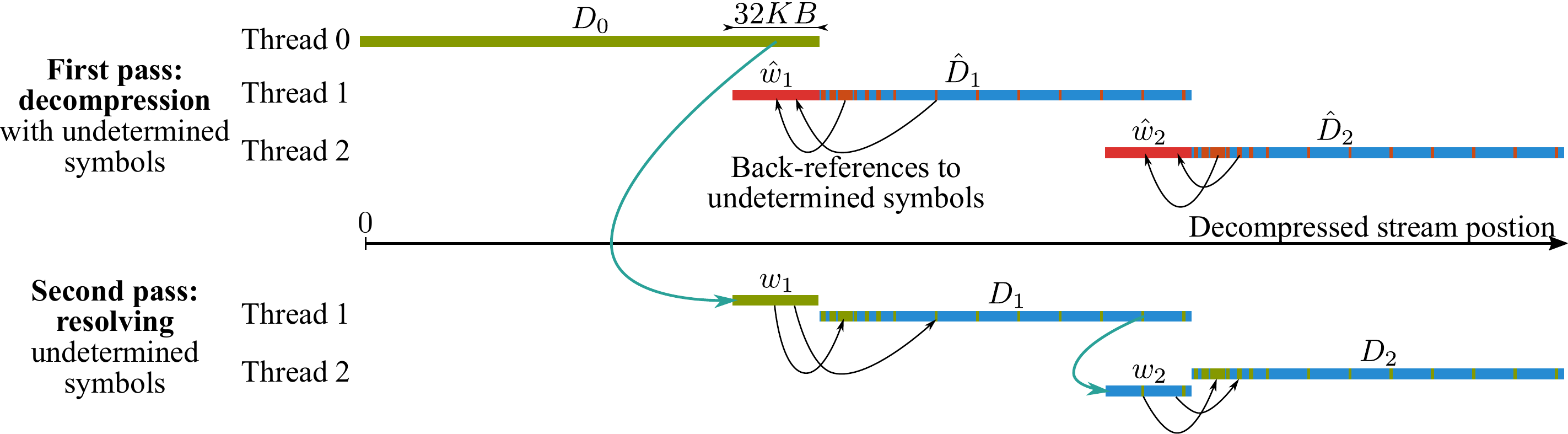}
\caption{Two passes decompression (\pugz). The first pass decompresses each part of the \gzip file in parallel with initially undetermined windows $\hat{w}$ containing unique symbols, for tracking back-references. The second pass resolves theses back-references with the initial context obtained from the previous part in the decompressed stream.}
\label{fig:pugz}
\end{figure*}

See Figure~\ref{fig:pugz} for an illustration.
While this scheme requires holding the whole decompressed stream in memory before translation, it enables the parallel decompression of any kind of data without relying on heuristics or prior knowledge of the file structure (unlike our FASTQ heuristic). The memory requirements can be reduced by processing in parallel only a portion of the file at a time.
A software implementation, created by modifying an existing highly-optimized single-threaded software (\libdeflate), is available at: \url{https://github.com/Piezoid/pugz}.

\section{Experimental Results}

We downloaded 192.8 GB of compressed FASTQ files (100 files) from the European Nucleotide Archive. The first file of each of the latest experiments posted in \url{ftp://ftp.sra.ebi.ac.uk/vol1/} as of April 2018 was selected, excluding files below 200 MB. We also excluded special blocked/multi-part \gzip files, as they are currently not handled by neither \libdeflate nor \pugz, but this is only due to an unimplemented special case in DEFLATE decompression. The list of files is available in our Github repository. 
Experiments were run on a 2x12-core 2.3 GHz Xeon E5-2670 v3 machine, 512 GB of RAM, NAS storage. 

\subsection{Success rate of decompressing at random location \label{sec:res-succrate}}

We investigated whether one could retrieve all sequences after decompressing at a random location in a compressed FASTQ file. This is a slightly easier flavor of Problem~\ref{prob:random}, as it focuses only on the DNA sequences inside a FASTQ file, disregarding headers and quality values. We first noticed that the compression level of a file plays a critical role. Therefore we partitioned our dataset into three compression levels, according to the UNIX '\texttt{file}' command: lowest (e.g. likely the result of \gzip $\texttt -1$),  normal (any parameter between \texttt{-2} and \texttt{-8}, usually \texttt{-6}), and highest (\gzip $\texttt -9$). Note that other \gzip-compatible compressors may report a compression level that does not match the performance of \gzip.

Table~\ref{tab:data} summarizes the results. At the lowest compression level, virtually no sequence after a sequence-resolved block contains undetermined characters. Only in two files (ERA966074 and ERA990245), a dozen of sequences out of respectively 11 and 27 million contain undetermined nucleotides. 
Therefore, one can perform virtually exact random accesses to low-compression files, requiring only around 52 MB of decompression to 'prime' the context. 

However at the normal and highest compression levels, only respectively 72\% and 37\% of sequences on average are fully determined after a sequence-resolved block. This is likely due to back-references that occur between DNA sequences and quality sequences or headers, which can also harbor DNA characters.  At the normal compression level, in 48\% of the files (data not shown), nearly all returned sequences (99.9--100\%) are unambiguous. For the rest of the files, either no sequence-resolved block is found or a variable fraction of sequences contain undetermined characters.

\begin{table}
\centering
\begin{tabular}{l|c|c|c|c}
\hline
\multicolumn{3}{c|}{\textbf{Dataset}} & \multicolumn{2}{c}{Random access to sequences}\\ \hline
 \shortstack{\strut Compress.\\level} & \shortstack{\strut Number\\of files} & \shortstack{Total size \\(GB)} & \shortstack{\strut Delay to sequence-\\resolved block (MB)} &  \shortstack{\strut Unambiguous\\sequences (\%)}\\ \hline
\textbf{Lowest} & 26 & 53.8  & 52.4 $\pm$ 55.8 & 100.0 $\pm$ 0.0\\ 
\textbf{Normal} & 68 & 111.8 & 387.5 $\pm$ 731.6 & 72.5 $\pm$ 37.6\\ 
\textbf{Highest} & 6 & 27.2  & 1,292.6 $\pm$ 1,531.9 & 36.8 $\pm$ 45.2\\ \hline
\textbf{Total}  &  100 & 192.8 &  317.8 $\pm$ 703.7  & 77.5 $\pm$ 36.5 \\ \hline
\end{tabular}
\caption{Compression level, number and size of files in our dataset. \textup{We performed random access decompression of sequences as per Section~\ref{sec:determine} at 4 different locations in each file: $\frac{1}{4}$th, $\frac{1}{3}$rd, $\frac{1}{2}$th, and $\frac{2}{3}$rd of the total file size, each time until the end of the file. "Delay to sequence-resolved block" reports the average number of bytes decompressed until a sequence-resolved block is found. "Unambiguous sequences" gives the percentage of sequences without any undetermined character returned by the heuristic parser after the first sequence-resolved block.}}
\label{tab:data}
\end{table}

\subsection{Propagation of initial contexts}

\begin{figure}
\centering
\includegraphics[width=.9\columnwidth]{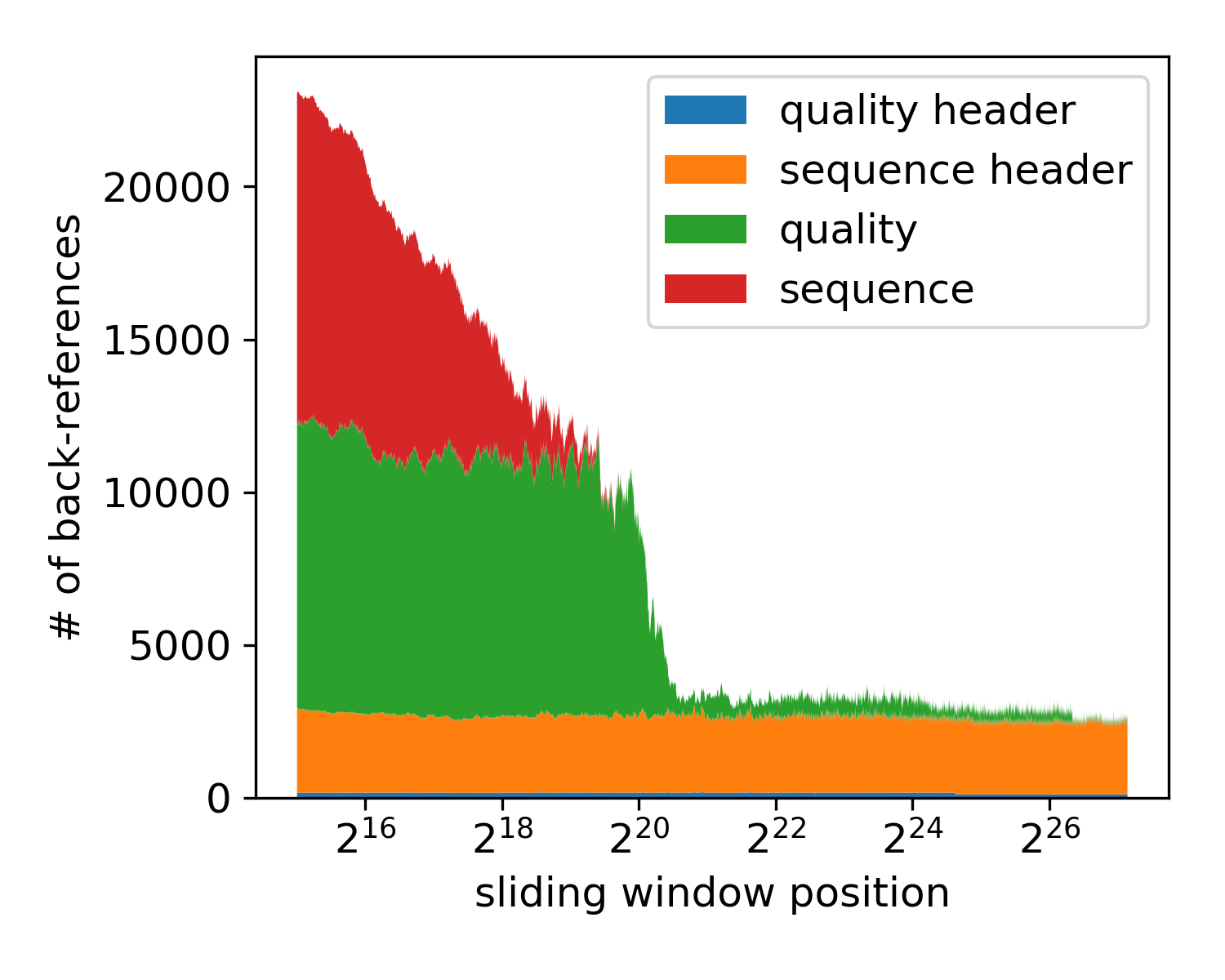}
\includegraphics[width=.9\columnwidth]{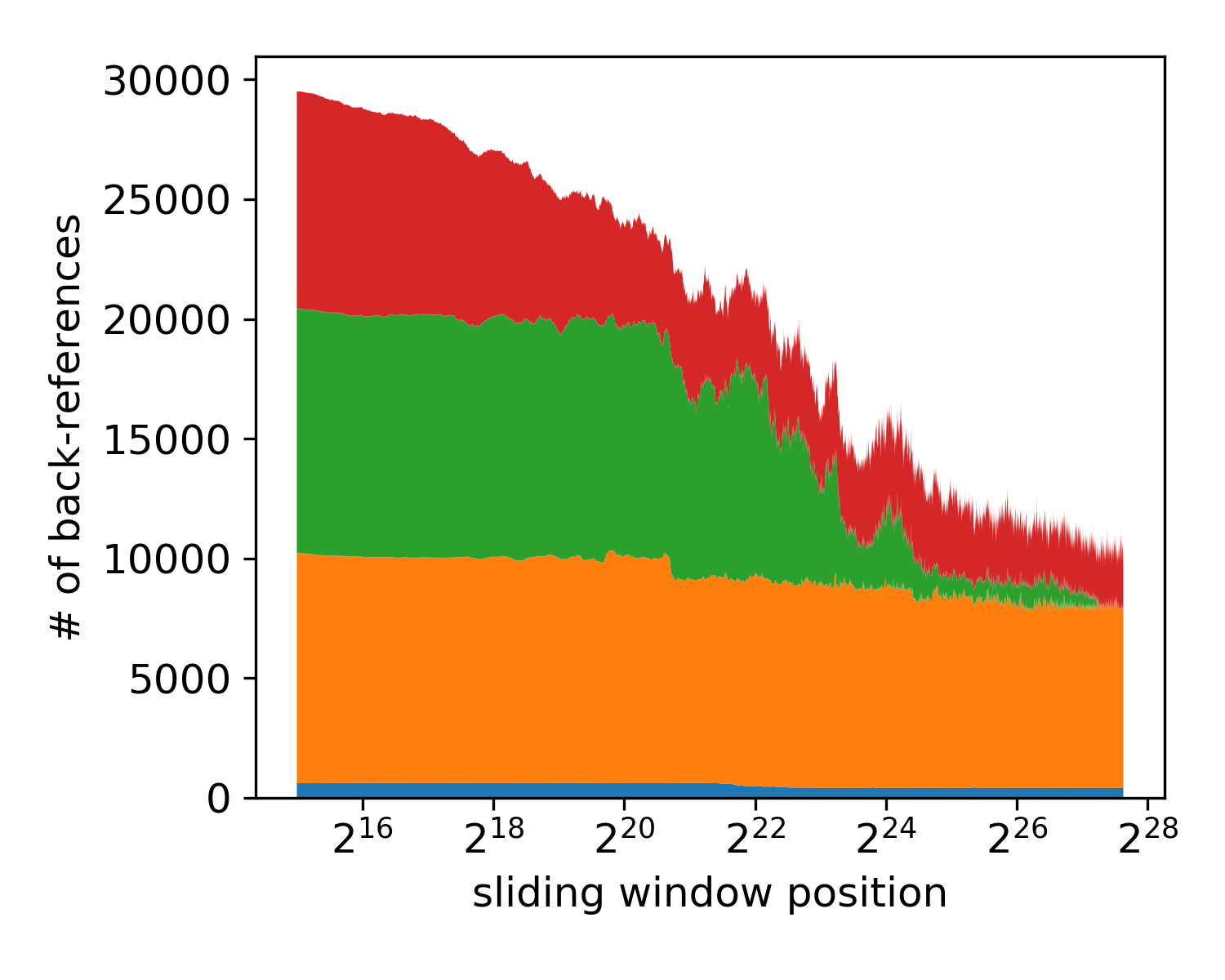}
\caption{Number and type of characters copied from an initial context during decompression of two \gzip-compressed FASTQ files, starting from a location inside the file. Characters are counted in sliding windows of size 32 KB. The analyzed files are (top) 
\texttt{DRC\_BKV\_01\_R1.fastq.gz} (normal compression) from ERA972077 decompressed at offset 160 MB,
and (bottom) \texttt{CNC\_CasAF3\_CRI1strepeat\_rep1\_R1.fastq.gz} (highest compression) from ERA987833, decompressed at offset 210 MB.}
\label{fig:porcelain}
\end{figure}

To further understand why some FASTQ files lend themselves to random access decompression and some do not, we instrumented our implementation to track how far characters from the initial undetermined context travel along matches. We also compared the undetermined context with the corresponding actual context and annotated each character by type: DNA, quality value, sequence header, or quality header (usually just the '+' character).
Figure~\ref{fig:porcelain} shows two instances of FASTQ files being decompressed from a random location. We observe that in the top plot (normal compression), none of the initial context characters that encode DNA sequences remain in matches after around position $2^{21}$ in the decoded stream; but a small amount of quality values do. Some headers characters remain until the end of the file. In other files with normal compression levels, 
none of the quality values from the initial context remain until the end of the file (data not shown). However, in the bottom plot (highest compression), parts of the DNA sequences remain in matches until the end of the file. This is likely due to \gzip trying harder at finding long/far matches instead of outputting literals.

\subsection{Parallel decompression speed}

We performed parallel decompression of three FASTQ files (of sizes 3--7.5 GB) at normal compression level. 
The files were preloaded into system memory to avoid IO bottlenecks: the purpose of this benchmark is to compare pure decompression speeds of \gunzip, \libdeflate, and \pugz. We ran  \pugz using 32 threads; \gunzip and \libdeflate cannot be multi-threaded.

We further observed that synchronizing outputs between threads, or piping to \texttt{wc}, degrades performance (10--20\%). Therefore we redirected all outputs to \texttt{/dev/null}. For \pugz, we allowed each thread to write to the output without synchronization, to mimic the behavior of a FASTQ parser (as in some applications, the order of the reads is irrelevant).

\begin{figure}[!htb]
    \centering
    \includegraphics[width=\columnwidth]{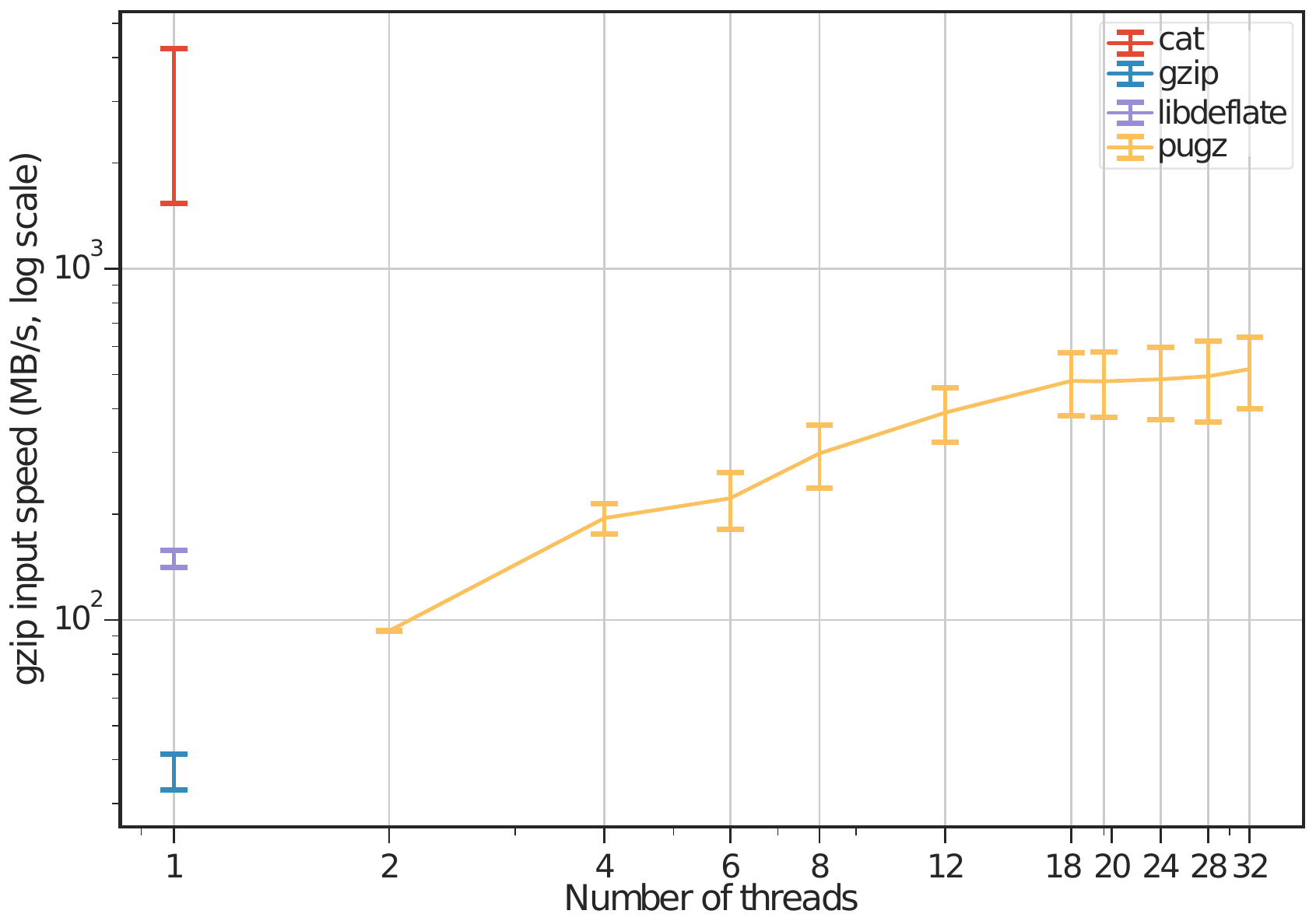}
\caption{Decompression speed of \pugz when executed with 2--32 threads compared to \gzip, \libdeflate, and the command \texttt{cat} as an upper bound. Lines show mean bandwidth while error bars represent standard deviation.}
\label{fig:scaling}
\end{figure}

Table~\ref{tab:res} shows that 
\pugz is 16.5x faster than \gunzip and 5.2x faster than \libdeflate. Figure~\ref{fig:scaling} shows parallel scaling performance.

\begin{table}
\centering
\begin{tabular}{|l|c|c|c|}
\hline

\textbf{Method} & \gunzip & \libdeflate & \pugz, 32 threads \\ \hline
\textbf{Speed} (MB/s) & 37 & 118 & 611 \\ \hline
\end{tabular}
\caption{Decompression speeds (megabyte of compressed data per second) for sequential and parallel \gzip-compatible software. 
\textup{The 3 first FASTQ files of experiments ERA970963, ERA973411 and ERA981545 were preloaded into memory and decompressed three times each. The average wall-clock time over compressed file size was recorded.}
}
\label{tab:res}
\end{table}

\section{Discussion}

We developed the first parallel decompression algorithm for \gzip-compressed ASCII files. Note that the current implementation requires the whole decompressed file to reside in memory, yet further engineering efforts could lift this limitation with little projected impact on performance. It also does not compute cyclic redundancy check (CRC32), nor handle multi-part \gzip files. Furthermore, it may be possible to decompress non-ASCII (binary) files, but this would require more careful determination of block boundaries.

We also demonstrated for the first time the decompression of sequences at random locations in \gzip-compressed FASTQ files. The method is near-exact at low compression levels, but often a large fraction of sequences contains undetermined characters at higher compression levels. This is due to long chains of back-references that trace back to the initial 32 KB context containing undetermined characters. It did not escape our attention that guessing those undetermined characters could be possible, but we did not yet explore this direction.

\bibliography{main}

\begin{thebibliography}{10}

\bibitem{abdelfattah2014gzip}
Mohamed~S Abdelfattah, Andrei Hagiescu, and Deshanand Singh.
\newblock Gzip on a chip: High performance lossless data compression on {FPGAs}
  using {OpenCL}.
\newblock In {\em Proceedings of the International Workshop on OpenCL 2013 \&
  2014}, page~4. ACM, 2014.

\bibitem{leon}
Ga{\"e}tan Benoit, Claire Lemaitre, Dominique Lavenier, Erwan Drezen, Thibault
  Dayris, Raluca Uricaru, and Guillaume Rizk.
\newblock Reference-free compression of high throughput sequencing data with a
  probabilistic de bruijn graph.
\newblock {\em BMC bioinformatics}, 16(1):288, 2015.

\bibitem{brown2011reconstructing}
Ralf~D Brown.
\newblock Reconstructing corrupt deflated files.
\newblock {\em Digital Investigation}, 8:S125--S131, 2011.

\bibitem{brown2013improved}
Ralf~D Brown.
\newblock Improved recovery and reconstruction of deflated files.
\newblock {\em Digital Investigation}, 10:S21--S29, 2013.

\bibitem{bwt}
Michael Burrows and D.~J. Wheeler.
\newblock A block-sorting lossless data compression algorithm.
\newblock Report 124, Digital Systems Research Center, Palo Alto, CA, USA, May
  1994.

\bibitem{deorowicz2013data}
Sebastian Deorowicz and Szymon Grabowski.
\newblock Data compression for sequencing data.
\newblock {\em Algorithms for Molecular Biology}, 8(1):25, 2013.

\bibitem{RFC1951}
L.~Peter Deutsch.
\newblock {DEFLATE} compressed data format specification version 1.3.
\newblock RFC 1951, May 1996.

\bibitem{RFC1950}
L.~Peter Deutsch and Jean-Loup Gailly.
\newblock {ZLIB} compressed data format specification version 3.3.
\newblock RFC 1950, May 1996.

\bibitem{nongreedy}
R.~Nigel Horspool.
\newblock The effect of non-greedy parsing in {Ziv-Lempel} compression methods.
\newblock In {\em Data Compression Conference, 1995. DCC'95. Proceedings},
  pages 302--311. IEEE, 1995.

\bibitem{Kreft2010}
Sebastian Kreft and Gonzalo Navarro.
\newblock {LZ77}-like compression with fast random access.
\newblock In {\em Proceedings of the 2010 Data Compression Conference}, DCC
  '10, pages 239--248, 2010.

\bibitem{li_2014}
Heng Li.
\newblock Random access to zlib-compressed files, 2014.
\newblock URL:
  \url{http://lh3.github.io/2014/07/05/random-access-to-zlib-compressed-files}.

\bibitem{samtools}
Heng Li, Bob Handsaker, Alec Wysoker, Tim Fennell, Jue Ruan, Nils Homer, Gabor
  Marth, Goncalo Abecasis, and Richard Durbin.
\newblock The sequence alignment/map format and {SAM}tools.
\newblock {\em Bioinformatics}, 25(16):2078--2079, 2009.

\bibitem{ryabko2005using}
B~Ya Ryabko and VA~Monarev.
\newblock Using information theory approach to randomness testing.
\newblock {\em Journal of Statistical Planning and Inference}, 133(1):95--110,
  2005.

\bibitem{sadakane2000improving}
Kunihiko Sadakane and Hiroshi Imai.
\newblock Improving the speed of {LZ77} compression by hashing and suffix
  sorting.
\newblock {\em IEICE transactions on fundamentals of electronics,
  communications and computer sciences}, 83(12):2689--2698, 2000.

\bibitem{sitaridi2016massively}
Evangelia Sitaridi, Rene Mueller, Tim Kaldewey, Guy Lohman, and Kenneth~A Ross.
\newblock Massively-parallel lossless data decompression.
\newblock In {\em Parallel Processing (ICPP), 2016 45th International
  Conference on}, pages 242--247. IEEE, 2016.

\bibitem{lz77}
Jacob Ziv and Abraham Lempel.
\newblock A universal algorithm for sequential data compression.
\newblock {\em IEEE Transactions on Information Theory}, 23(3):337--343, 1977.

\end{thebibliography}
\section{Acknowledgements}

This work was supported by an Inria ADT grant (SeedLib) and the INCEPTION project (PIA/ANR-16-CONV-0005).

\section{Appendix}

\subsection{DEFLATE block decompression checks \label{appendix:block}}

\begin{itemize}
\item 
The first bit of the block needs to be $0$, indicating that it is not the last block in the stream.  This implies that we will never seek to the very last block.
\item
The next two bits encoded the block type, but there are only three valid types: the remaining 2-bit code is invalid.
\item
The next bits may contain a dynamic Huffman tree, where the encoding may be invalid in different ways.
\item
A decompressed block may only contain valid ASCII characters.
\item Invalid back-references with an offset exceeding the context size (32 KiB)
\item Decompressed block should be larger than 1 KiB and smaller than 4 MiB.
\end{itemize}

\subsection{Heuristic extraction of DNA sequences in DEFLATE block \label{appendix:heuristic}}

Given a decompressed block, the procedure returns all maximal non-overlapping substrings that match the following grammar: $TD^+(U^+D^+)^*T$, where $T$ is a newline or undetermined character, $D$ is a nucleotide (A,C,T,G,N), $U$ is an undetermined character, and the notations $X^+$ (resp. $X^*$) classically represents a non-empty (resp. possibly empty) sequence of consecutive $X$'s.

The leading and trailing $T$ characters of each sequence are removed from the results, but are needed to filter out parts of quality sequences that look like DNA. In the implementation, matches shorter than a minimum read length are discarded, and a special case is needed to handle sequences that span two blocks.

Some of the false positive sequences returned by the parser can be trivially detected and removed on files where reads all have the same lengths. However, due to possible matches to a DNA sequence from quality sequences or even headers (both of which often contain undetermined characters), a sequence may have undetermined characters even after a sequence-resolved block. 

\end{document}